\documentclass[aps,shownopacs,superscriptaddress,nofootinbib,preprint,11pt]{revtex4}
\usepackage[section,subsection,subsubsection]{extraplaceins}
\usepackage{hyperref}
\usepackage{amsmath}
\usepackage{float}
\usepackage{graphics}
\usepackage{graphicx}
\usepackage{epsfig}
\usepackage{psfrag}
\usepackage{color}
\usepackage{slashed}

\usepackage{amsfonts}
\usepackage{amssymb}

\newcommand*{\be}{\begin{equation}}
\newcommand*{\ee}{\end{equation}}
\newcommand*{\bea}{\begin{eqnarray}}
\newcommand*{\eea}{\end{eqnarray}}

\newcommand{\comment}[1]{}

\newcommand{\cref}[1]{Chapter~\ref{c.#1}}

\def\beq{\begin{equation}}
\def\eeq{\end{equation}}
\def\bea{\begin{eqnarray}}
\def\eea{\end{eqnarray}}
\def\ba{\begin{array}}
\def\ea{\end{array}}
\def\bi{\begin{itemize}}
\def\ei{\end{itemize}}
\def\be{\begin{enumerate}}
\def\ee{\end{enumerate}}
\def\bc{\begin{center}}
\def\ec{\end{center}}
\def\bt{\begin{table}}
\def\et{\end{table}}
\def\btb{\begin{tabular}}
\def\etb{\end{tabular}}

\def\lsim{\raise0.3ex\hbox{$\;<$\kern-0.75em\raise-1.1ex\hbox{$\sim\;$}}}
\def\gsim{\raise0.3ex\hbox{$\;>$\kern-0.75em\raise-1.1ex\hbox{$\sim\;$}}}

\def\beq {\begin{equation}}
\def\eeq {\end{equation}}
\def\bea {\begin{eqnarray}}
\def\eea {\end{eqnarray}}

 \preprint{TIFR/TH/15-12}
\begin{document}

\title{Looking for lepton flavour violation in supersymmetry at the LHC}
\author{Monoranjan Guchait}
\email{guchait@tifr.res.in}
\affiliation{Department of High Energy Physics, Tata Institute of Fundamental Research, Homi Bhabha Road, Colaba, Mumbai 400 005, India}

\author{ Abhishek M. Iyer}
\email{abhishek@theory.tifr.res.in}
\affiliation{Department of Theoretical Physics, Tata Institute of Fundamental Research, Homi Bhabha Road, Colaba, Mumbai 400 005, India}

\author{Rickmoy Samanta}
\email{rickmoy@theory.tifr.res.in}
\affiliation{Department of Theoretical Physics, Tata Institute of Fundamental Research, Homi Bhabha Road, Colaba, Mumbai 400 005, India}

\begin{abstract}
 We consider models of supersymmetry which can incorporate sizeable mixing between different generations of sfermions.
 While the mixing is constrained by the non-observation of various flavour changing neutral current (FCNC) processes, there exist regions of SUSY parameter space where the effects of such mixing can be probed at colliders. In this work, we explore this possibility by focussing on the slepton sector. The sleptons are produced through cascade decays in direct neutralino-chargino ($\chi_2^0\chi_1^\pm$) pair production at the LHC. The final state is characterized by 3 leptons and missing energy. We probe the lepton flavour violating (LFV) vertex originating from $\chi_2^0$ decay and identify a distinct and unambiguous combination of the tri-lepton final state which include a lepton pair with same flavour and same sign (SFSS) in addition to a pair with opposite flavour and opposite sign (OFOS). This combination of tri-lepton final state containing both OFOS and SFSS pairs, can not only suppress the SM background but can also differentiate the flavour violating decays of $\chi_2^0$ from its corresponding flavour conserving decays. We present results for various signal benchmark points taking into account background contributions assuming two luminosity options (100 fb$^{-1}$ and 1000 fb$^{-1}$) for LHC Run 2 experiment.

\end{abstract}
\vskip .5 true cm

\maketitle
\section{Introduction}

The experiments dedicated towards the investigation of flavour physics are considered to be one of best indirect ways to establish the existence of new physics (NP). 
They play an important role in constraining the viability of various new physics scenarios, thereby complementing the direct collider searches.
The effects which give rise to large flavour changing neutral currents (FCNC), can also be potentially probed at the colliders. For instance, the possibility of observing a flavour violating Higgs decay at the Large Hadron Collider  (LHC) was discussed in \cite{DiazCruz:1999xe,Blankenburg:2012ex,Harnik:2012pb}. 
Further, an observation of a  $2.5~\sigma$ excess in the $H\rightarrow \tau\mu$ channel by CMS \cite{CMS:2014hha} in the LHC experiment has generated a lot of interest in this sector and has led to a plethora of analysis \cite{Pilaftsis:1992st,Brignole:2003iv,Arganda:2004bz,Azatov:2009na,Arhrib:2012mg,Falkowski:2013jya,Arana-Catania:2013xma,Arroyo:2013kaa,Arganda:2014dta,Sierra:2014nqa,Heeck:2014qea,Lee:2014rba,Crivellin:2015mga,deLima:2015pqa}. 
The leptonic sector in the Standard Model (SM) is also interesting owing to the absence of FCNC. This can be attributed to the massless nature of neutrinos in the SM. The observation of neutrino oscillations, which consequently led to a confirmation of the massive nature of left handed neutrinos,  resulting in a non-zero decay rate for rare processes like $\mu\rightarrow e\gamma$. The predicted branching ratio (BR) in the SM, however is negligibly small ($\sim 10^{-40}$) due to the tiny neutrino mass  and is beyond the sensitivity of the current flavour experiments. There exist several extensions of the SM which contribute to rare processes such as $\mu\rightarrow e\gamma$ via loops, enhancing the BR substantially to $\sim 10^{-13}-10^{-15}$ and expected to be within the reach of the indirect flavour probes. Needless to say,  an observation of such processes is a definitive signal of the presence of physics beyond the SM. Therefore, looking for a signal of lepton flavour violation (LFV)  directly or indirectly is a challenging avenue to find NP.
Following this argument, we explore the possibility of observing lepton flavour violation at the LHC.

There are several models in literature which discuss the possibility of flavour violation in the leptonic sector.
 In the current analysis we focus on the supersymmetric extensions of the SM which can possess soft masses having significant flavour mixing in the mass basis of fermions. This can lead to new contributions to the BR of rare processes.
For instance, soft masses with flavour mixing can arise in see-saw extensions of SUSY \cite{Borzumati:1986qx,Hall:1985dx,Masiero:2004js} and also inspired by SUSY GUT \cite{Masiero:2002jn,Masiero:2005ua,Calibbi:2006nq,Hirsch:2012ti,Calibbi:2012gr}. Alternatively, introduction of flavour symmetries \cite{Feng:2007ke,Feng:2009bd}, models with messenger matter mixing in gauge mediated supersymmetry breaking (GMSB)\cite{Fuks:2008ab,Shadmi:2011hs,Abdullah:2012tq,Calibbi:2013mka,Calibbi:2014yha}, models with R-symmetric supersymmetry \cite{Kribs:2007ac,Kribs:2009zy}, supersymmetric theories in the presence of extra-spatial dimensions \cite{Nomura:2007ap,Nomura:2008gg,Nomura:2008pt} \textit{etc.} also lead to flavourful soft masses. Scenarios in which mass splitting lead to flavour violation have been considered in \cite{Galon:2013jba,Abdullah:2012tq,Calibbi:2014pza}. Such extensions in general lead to flavoured soft masses and depending on the parameters can lead to observable rates for the flavour violating decays in the squark and leptonic sector.

Flavour mixing in the sfermion mass matrices can be probed at the collider by the flavour violating decay of a sparticle of flavour (say $i$) into a fermion of flavour $j$ where $j\neq i$. Flavour violating decays of sleptons were studied in the context of $e^+e^-$ linear collider \cite{Krasnikov:1994hr,Krasnikov:1995qq,ArkaniHamed:1996au,Hisano:1998wn,Guchait:2001us,Carquin:2011rg,Abada:2012re}. In Ref. \cite{ArkaniHamed:1997km}
the authors studied the possibility of observing CP violation from slepton oscillations at the LHC and NLC. At the LHC, the sleptons can be produced either through Drell-Yan (DY) process or by cascade decays from heavier sparticles. Subsequent flavour violating decays of sleptons produced by DY
were studied in \cite{Bityukov:1997ck,Krasnikov:1996np} while those produced by cascade decays were studied in \cite{Agashe:1999bm,Hinchliffe:2000np,Hisano:2002iy,Kitano:2008en,Kaneko:2008re,Andreev:2006sd,Deppisch:2007rm}. Probing LFV through the measurement of splitting in the mass eigenstates of sleptons was considered in \cite{Allanach:2008ib,Buras:2009sg,Galon:2011wh}.
In this paper we report on our study of flavour violation in the leptonic sector by producing sleptons in cascade decays through pair production of neutralino-chargino at the future LHC experiments.

Starting with MSSM, we write the most general structure for the slepton mass matrix.  The constraints on the model from the non-observation of flavour violating processes can be expressed by working in the mass-insertion approximation (MIA) \cite{Hall:1985dx,Gabbiani:1988rb} in terms of bounds on the flavour violating parameter $\delta_{ij,}~i\neq j$ as defined in Eq.\ref{delta} \cite{Gabbiani:1996hi}. A non-zero $\delta_{ij}$ also opens up the possibility of flavour violating decay as far as collider implications of flavoured slepton masses are concerned.

Our goal is to probe the flavour violating decay in the case of first two generations in the slepton sector in SUSY. In this context strong bounds exist on the flavour violating parameter, coming primarily from the non-observation of $\mu\rightarrow e\gamma$ \cite{Adam:2013mnn}.
There exist regions of parameter space where these bounds can be relaxed owing to cancellations between different diagrams contributing to this process, thereby giving access to probe LFV at the colliders.

In this letter we explore this possibility to look for LFV decays considering neutralino-chargino pair production in proton-proton collisions,  which eventually leads to three lepton and missing energy final state. 
The tri-lepton final state is characterized  by the presence of two leptons with opposite flavour and opposite sign combination (OFOS).
The presence of LFV in the tri-lepton final state is ensured by demanding a combination of same flavour same sign (SFSS) lepton pair along with the OFOS combination.
 While an imposition of this SFSS criteria along with OFOS has a tendency to decrease the signal, it aids in suppressing  the backgrounds due to SM and SUSY significantly.

The paper is organized as follows: In Section \ref{model} we discuss the model set-up introducing the various parameters relevant for the analysis in the framework of a simplified model. Relevant regions of parameter space consistent with the flavour constraints and conducive to be probed at the colliders are identified in this section. In Section \ref{sigbg} we explain our choice of OFOS and SFSS combination to extract the signal with a detailed description of the simulation. The results of the simulation for the background and the representative points for the signal events are presented. 
In Section \ref{lhc} we show regions of the parameter space which can be probed at the LHC Run 2 experiment in the near future.
 We conclude in Section \ref{conclusion}.
\section { Model Parametrization}
\label{model}
In this section  we introduce the basic model set-up and related parameters necessary to describe LFV. In order to reduce the dependence on many parameters, we consider a simplified SUSY model (SMS) approach with only left handed sleptons, wino and a bino while decoupling the rest of the spectrum. The $\mu$ term is assumed to be $\sim 1$ TeV to make the neutralino/chargino dominantly composed of gauginos with a very small higgsino component. In this case, the mass of $\chi_2^0$, the second lightest neutralino and $\chi_1^\pm$, the lightest chargino, are roughly the same as $\sim M_2$, the mass of the $SU(2)$ gauginos. The lightest neutralino $\chi_1^0$, which is assumed to be the lightest supersymmetric particle (LSP)  has mass $\sim M_1$, same as the mass  of the $U(1)$ gaugino.
 
 For the slepton sector we focus on the flavour violation in the left handed sector making the right handed sleptons very heavy and set the left-right chiral mixing in the slepton mass matrix to be negligible. 
For simplicity, we assume only two generations. With these assumptions, the left handed slepton mass matrix in the basis $l_F\equiv(\tilde e_F, \tilde \mu_F)$ is given as,
\begin{equation}
\tilde m^2=\begin{bmatrix}
m^2_{L_{11}}&m^2_{L_{12}}\\
m^2_{L_{12}}&m^2_{L_{22}}
\end{bmatrix},
\label{sleptonmassmatrix}
\end{equation}
where $F$ denotes the flavour basis (SUPER CKM) for the sleptons.
In this basis the flavour violating parameter $\delta_{12}$ is parametrised as \cite{Hall:1985dx,Gabbiani:1988rb},
\begin{equation} 
\delta_{12}=\frac{m^2_{L_{12}}}{\sqrt{m^2_{L_{11}}m^2_{L_{22}}}}.
\label{delta}
\end{equation}
Naturally, this flavour violating parameter  $\delta_{12}$ is coupled to the rates corresponding to flavour violating rare decays in the first and second generation lepton sector. Hence an upper bound on this parameter exists due to non-observations of these rare decays like $\mu\rightarrow e\gamma$ \cite{Adam:2013mnn}, $\mu-e$ conversion \cite{Wintz:1996va} and $\mu\rightarrow eee$ \cite{DeGerone:2011fg}.

In order to obtain the mass eigenvalues of the sleptons, the matrix in Eq.\ref{sleptonmassmatrix} can be rotated into a diagonal form by an angle $\theta$ given by,
\begin{equation}
\sin2\theta=\frac{2m^2_{L_{12}}}{m^2_{L_2}-m^2_{L_1}},
\end{equation}
where $m^2_{L_i}$ are the eigenvalues.
It can be related to the flavour violating parameter  $\delta_{12}$ as,
\begin{equation}
\delta_{12}=\frac{\sin2\theta(m^2_{L_2}-m^2_{L_1})}{2m_L^2}
\label{FV}
\end{equation}
where $m_L=\frac{m_{L_1}+m_{L_2}}{2}$.
The structure of the mass matrix, Eq.\ref{sleptonmassmatrix} allows for the possibility of flavour oscillations similar to neutrino flavour oscillations. The probability $P(\tilde e_F\rightarrow \mu)$ of a flavour eigenstate $\tilde e_F$ decaying into a muon is given by \cite{ArkaniHamed:1996au},
\begin{eqnarray}
P(\tilde e_F\rightarrow \mu) &=& \sin^22\theta\frac{(\Delta m^2)^2}{4\bar m^2\Gamma^2+(\Delta m^2)^2}BR(\tilde \mu \rightarrow \mu),\nonumber\\
&\sim& \sin^22\theta ~ BR(\tilde \mu \rightarrow \mu)~\text{for}~ \Gamma\ll \Delta m^2,
\end{eqnarray}
with $\Delta m^2=m_{L_2}^2-m_{L_1}^2$.
The above expression can be re-expressed in terms of the parameter $\delta_{12}$ from Eq.\ref{FV}. Thus the branching ratio for the flavour violating decay, $\chi^{0}_2\rightarrow e~\tilde e\rightarrow e~\mu~\chi^{0}_1$ can be computed as, 
\begin{equation}
BR(\chi^{0}_2\rightarrow e~\mu~\chi^{0}_1)=\mathcal{B}_{LFV}~BR(\chi_2^{0}\rightarrow \tilde e~e)~BR(\tilde e\rightarrow e \chi_1^{0})+e\leftrightarrow\mu
\label{chi2decay}
\end{equation}
Here the suppression factor due to flavour violation is given by,
\begin{equation}
\mathcal{B}_{LFV}=\sin^2 2\theta=\left(\frac{m_L~\delta_{12}}{\Delta m_{12}}\right)^2,
\label{blfv}
\end{equation}
where $\Delta m_{12}=m_{L_2}-m_{L_1}$.

As mentioned before, bounds on $\delta_{12}$ and hence $\mathcal{B}_{LFV}$ can be obtained by taking into account the experimental upper limit on the  $BR(\mu\rightarrow e\gamma) < 5.7\times 10^{-13}$ \cite{Adam:2013mnn}. The higher dimensional operator contributing to this process is parametrized as \cite{Calibbi:2015kja},
\begin{equation}
\mathcal{L}_{FV}=e\frac{m_l}{2}~\bar{e}~\sigma_{\alpha\beta}\left(A_LP_L+A_RP_R\right)~\mu~ F^{\alpha\beta},
\end{equation}
 where the model dependence is captured by the Wilson coefficients $A_{L,R}$. 
The branching ratio for this process is then given by \cite{Calibbi:2015kja},
\begin{eqnarray}
BR(\mu\rightarrow e\gamma)=\frac{48 \pi^3}{G_F^2}\left(|A_L|^2+|A_R|^2\right).
\label{meg}
\end{eqnarray}
In our considered model, $A_R\equiv 0$, as the right handed sleptons are assumed to be very heavy.
$A_L$ on the other hand receives three contributions due to chargino, neutralino and bino mediated diagrams and is given as \cite{Calibbi:2015kja},
\begin{eqnarray}
A_L=\frac{\delta_{12}}{m_L^2}\left(\frac{\alpha_Y}{4\pi}f_n\left(\frac{M_1^2}{m_L^2}\right)+\frac{\alpha_Y}{4\pi}f_n\left(\frac{M_1^2}{m_L^2}\right)+\frac{\alpha_2}{4\pi}f_c\left(\frac{M_2^2}{m_L^2}\right)\right)
\end{eqnarray}
where $f_{n,c}$ are loop factors defined in \cite{Calibbi:2015kja} with a non-trivial mass dependence of related sparticles and $\alpha_Y,\alpha_2$ are the $U(1)_Y$ and $SU(2)$ gauge couplings.
	
\begin{figure}
	\begin{tabular}{cc}
	\includegraphics[width=8cm]{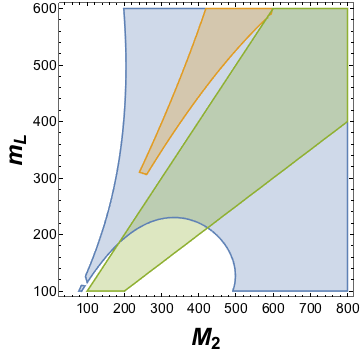}&\includegraphics[width=8cm]{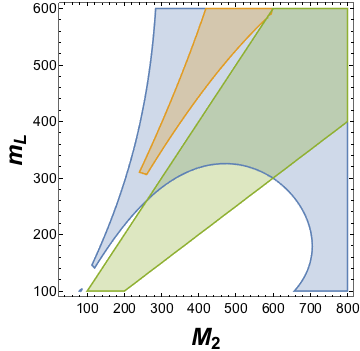}
\end{tabular}
	\caption{ Region satisfying Eq.\ref{parametrization1} and \ref{parametrization2} (green), while the orange regions satisfy the $\mu \rightarrow e\gamma$ constraint for $\delta_{12}=0.1$.
		The blue regions are allowed by the upper bound on BR($\mu \rightarrow e\gamma$) for $\delta_{12}=0.01$ (left) and $\delta_{12} =0.02$ (right). Units of mass are in GeV.}
	\label{parameterplot}
\end{figure}

 The analysis can be simplified again by choosing the following parametrization for the mass $M_1$ of the (LSP) $\chi_1^0$,
\begin{equation}
M_1=\frac{M_2}{2},
\label{parametrization1}
\end{equation}
which is the relation at the electroweak scale due to unification of gaugino masses at the GUT scale. For the sleptons we choose,
\begin{equation}
M_2>m_L>M_1.
\label{parametrization2}
\end{equation}
This relation assumes that the intermediate sleptons in $\chi_2^0$ decay are produced on-shell by requiring that they are lighter than the mass of $\chi^{0}_2$ $\simeq M_2$. 
Under these assumptions, we try to find the available range of parameters allowed by existing $\mu\rightarrow e\gamma$ constraints as will be discussed later.
Fig.\ref{parameterplot} shows the region in the $M_2-m_L$ plane for which the conditions in Eq.\ref{parametrization1}~and~\ref{parametrization2}~are satisfied (green region). It depicts the region of parameter space which is of interest as far as collider implications are concerned as discussed in this paper.

 The blue region shows the parameter space for which $BR(\mu\rightarrow e\gamma) <5.7 \times 10^{-13}$ is satisfied for $\delta_{12}=0.01$ in the left plot and for $\delta_{12}=0.02$ in the right plot. As expected, due to the smaller value of $\delta_{12}$, the blue region in the left plot has a larger overlap with the green region as compared to the plot in the right, thereby admitting smaller slepton masses.
  The orange region in both the plots shows the parameter space for which $BR(\mu\rightarrow e\gamma) <5.7 \times 10^{-13}$ is satisfied for $\delta_{12}=0.1$.
  We find that there is virtually no overlap with the region which is of interest to us from the view of collider searches.

\begin{figure}
	\begin{tabular}{cc}
	\includegraphics[width=8cm]{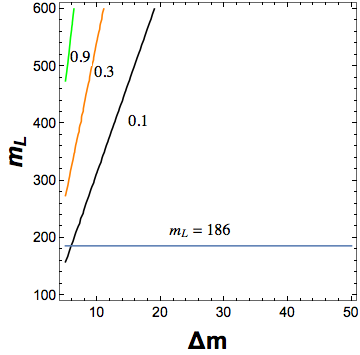}&\includegraphics[width=8cm]{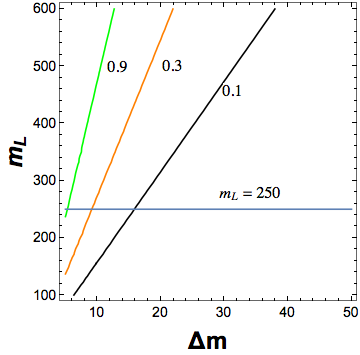}
	\end{tabular}
	\caption{Contours of $\mathcal{B}_{LFV}$  for $\delta_{12}$=0.01 (left) and $\delta_{12}=0.02$ (right). The horizontal blue line is excluded by $BR(\mu\rightarrow e\gamma)$ for $\delta_{12}=0.01$ (left) and $\delta_{12} =0.02$ (right). The units of mass are in GeV.}
	\label{deltammL}
\end{figure}

It would be interesting to estimate the  suppression factor $\mathcal{B}_{LFV}$ corresponding to the allowed region in the $M_2-m_L$ plane for the  values of $\delta_{12}$ in Fig.\ref{parameterplot}.
As seen in Eq.\ref{blfv}, the parameter $\mathcal{B}_{LFV}$,  which determines the rate for LFV, is sensitive to the mass-splitting $\Delta m =m_{L_2}-m_{L_1}$ and $m_L$. $\mathcal{B}_{LFV}$ increases with $\delta_{12}$ which can only be accommodated with a larger $m_L$. Thus smaller values of $\delta_{12}$ are not conducive to generate a large $\mathcal{B}_{LFV}$. $\mathcal{B}_{LFV}$ is also inversely proportional to the mass splitting $\Delta m$. However, it cannot increase indefinitely as $\mathcal{B}_{LFV}\leq 1$, leading to a lower bound on $\Delta m$.
Fig.\ref{deltammL} demonstrates the contours of constant $\mathcal{B}_{LFV}$  in the $\Delta m-m_L$ plane. We find that for $\delta_{12}=0.02$,  slepton in excess of 250 GeV are required to get $\mathcal{B}_{LFV}\geq 0.1$, while being consistent with the flavour constraints (overlap of blue and green region) in Fig. \ref{parameterplot}. 

\section{Signal and Background simulations}
\FloatBarrier
\label{sigbg}
As mentioned in the introduction, we probe the signal of LFV in slepton decay producing it via cascade decays of sparticles which are produced in proton-proton collisions at the LHC. Here we focus on $\chi_1^\pm\chi_2^0$ production which eventually leads to  a tri-lepton final state as, 
\begin{eqnarray}
pp\rightarrow\begin{cases}
\chi_2^{0}\rightarrow l_i^\pm \tilde l_i^\mp\rightarrow l_i^\pm  l_j^\mp\chi_1^0,\;\;\; i\neq j,\\
\chi_1^\pm\rightarrow l_i^\pm\nu \chi_1^0,       
\end{cases}
\label{decay}              
\end{eqnarray}
where $i,j$ denote flavour indices ($e,\mu$). 
The flavour violating vertex causes the decay of a slepton ($\tilde l_i$), coming from $\chi_2^0$ decay, in Eq.\ref{decay}, into a lepton of flavour $l_j$ with $i\neq j$.
It is clear from the above process that the signature of LFV  is the presence of 3 leptons of which 2 leptons are with opposite flavour and opposite sign (OFOS) in addition to missing energy ($\slashed E$) due to the presence of two LSP and neutrino. The leptons with OFOS originate from  $\chi^{0}_2$ decay while the third lepton comes from the $\chi^\pm_1$ decay. 
Thus, following this decay scenario, it is possible to have 8 combinations of tri-leptons, each having at least one OFOS lepton pair as, \\
\begin{eqnarray}
e^{+} e^{+} \mu^{-}~;e^{-} e^{-} \mu^{+}~;\mu^{-} e^{+} \mu^{-}~;\mu^{+} e^{-} \mu^{+} \nonumber\\
e^{+} e^{-} \mu^{+}~;e^{-} e^{+} \mu^{-}~;\mu^+ e^{+} \mu^{-}~; \mu^{-} e^{-} \mu^{+}. 
\label{combinations}
\end{eqnarray}

On the other hand, the pair production of $\chi^\pm_1 \chi_2^0$ will also give rise to tri-lepton final state with a flavour conserving decay of $\chi_2^0$ \textit{i.e.} $\chi_2^0\rightarrow l^+l^-\chi_1^0$. Note that this flavour conserving decay scenario also results in 8 combinations of tri-lepton final state given as
\begin{eqnarray}
e^{+} \mu^{+} \mu^{-}~;e^- \mu^{+} \mu^{-}~;\mu^{-} e^{+} e^{-}~;\mu^{+} e^{+} e^{-} \nonumber\\
\mu^{+} \mu^{+} \mu^{-}~;e^{-} e^{+} e^{-}~;e^{+} e^{+} e^{-}~; \mu^{-} \mu^{-} \mu^{+}
\label{fccombinations}
\end{eqnarray}
out of which four combinations of OFOS exist as seen in the first line of Eq.\ref{fccombinations}. It is clearly a potential background corresponding to the signal channel in Eq.\ref{combinations} and expected to have the same rate as signal. However, a closer look at these two final states in Eq.\ref{combinations} and \ref{fccombinations} reveals a characteristic feature. For example, in the case of signal, out of the 8 combinations of tri-leptons with OFOS combinations, notice that four combinations shown in the first line in Eq.\ref{combinations}, also possess a pair of leptons with same flavour and same sign (SFSS) which are absent in the background final states, shown in Eq.\ref{fccombinations}.
The rest of the states with OFOS combination in Eq.\ref{combinations} are identical to the final states given in Eq.\ref{fccombinations}. We exploit this characteristic feature to extract the LFV signal events out of all three lepton events including all backgrounds. Thus our signal is composed of three leptons having combinations of both OFOS and SFSS together, which is an unambiguous and robust  signature of LFV in SUSY.
Note that while choosing a clean signature of  LFV decay in SUSY, we pay a price by a factor of half as is clear from Eq.\ref{combinations}. However, this specific choice of combinations in tri-leptons is very powerful in eliminating much of the dominant SM backgrounds arising from $WZ$ and $t \bar t $ following leptonic decays of $W/Z$ and top quarks.
 
 We now discuss our simulation strategy to estimate the signal rates while suppressing the SM and SUSY backgrounds.
We performed simulations for both signal and background using  
{\tt{ PYTHIA8}} \cite{Sjostrand:2007gs} at 14 TeV centre of mass energy and applying the following selections: \\
$\bullet$~\textbf{Jet selection:} The jets are reconstructed using {\tt{FastJet}} \cite{Cacciari:2011ma} and based on anti 
$k_T$ algorithm \cite{Cacciari:2008gp} setting the jet size  parameter $R=0.5$.
The jets passing the cuts on transverse momentum $p_T^j \ge 30{\rm ~GeV}$, pseudo-rapidity $|\eta^j|\le 3.0,
\label{eq:jcut}
$ are accepted.\\
$\bullet$~\textbf{Lepton selection:}
Our signal event is composed of three leptons and are selected according to the following requirement on their transverse momenta and the pseudo-rapidity: 
$p_{T}^{\ell_{1,2,3}} \ge 20,20,10 {\rm~GeV}; ~|\eta^{\ell_{1,2,3}}| \le 2.5,
\label{eq:l1cut}$
where the leptons are $p_T$ ordered with $p_T^{\ell_1}$ being the hardest one. In addition, the leptons are also required to be isolated \textit{i.e.} free from nearby hadronic activities.
It is ensured by requiring the total accompanying transverse energy, which is the scalar sum of  transverse momenta of  jets within a cone of size $\Delta R(l,j) \le 0.3 $ around the lepton, is less than 10$\%$  of the transverse momentum  of the corresponding lepton. \\
$\bullet$~\textbf{Missing transverse momentum:} We compute the missing transverse momentum by carrying out a vector sum over the momenta of all visible particles and then reverse its sign. Since $\slashed{p_T}$ is hard in signal events, so
we apply a cut $\slashed{p_T}\geq100 $ GeV. \\
$\bullet$~\textbf{Z mass veto:} We require that in three lepton events, the invariant mass of two leptons with opposite sign and same flavour should  not lie in the mass window
$m_{ll} = M_Z\pm 20$ GeV. It helps to get rid of significant amount of $WZ$ background.\\
$\bullet$~\textbf{$b$ like jet selection:} The $b$ jets are identified through jet-quark matching \textit{i.e.} those jets which  lie with in $\Delta R(b,j)<0.3$ are assumed to be $b$ like jets.\\
$\bullet$~\textbf{OFOS:} Our signal event is characterised by the requirement that it has at least one lepton pair with opposite flavour and opposite sign.\\
$\bullet$~\textbf{SFSS:} We require the presence of SFSS combination along with OFOS combination in three lepton final state, which is the characteristic of our signal. As stated before, this criteria is very effective in isolating the background due to the same SUSY process but for their subsequent flavour conserving decays, in particular for $\chi_2^0$ decay.

We perform our analysis by choosing various representative points in the SUSY parameter space. 
The spectrum is generated using {\tt{ SUSPECT}} \cite{Djouadi:2002ze} and the decays of the sparticles are computed using {\tt{SUSYHIT}} \cite{Djouadi:2006bz}. Table \ref{spectrum} presents the six representative points (A-F) for which we discuss the details of our simulation. From A to F, the spectrum is characterized by increasing masses of gauginos, with the slepton mass $m_L$ lying midway between the two, $m_L=\left(M_1+M_2\right)/2$. 


\begin{table}[here]
	\begin{center}
		\begin{tabular}{|c|cccccc|}
			\hline
			Spectrum Characteristics                                        &A    & B   & C     &D     &E    &F      \\  
			\hline 
			$\chi_{2}^{0}/\chi^\pm_1$                             &210  &314  &417    &518   &619  &718                           \\
			$\chi_{1}^0$                             &95.8 &144  &193    &241   &290  &339                         \\
			$m_L$                                &156  &229  &303    &377   &452  &526                    \\
			$BR(\chi^{0}_2 \rightarrow  \tilde e_{L} e) $     &0.13 &0.15 &0.16   &0.16    &0.16  &0.16                            \\
			$BR(\chi^{0}_2 \rightarrow  \tilde\mu_{L} \mu)$  &0.13 &0.15 &0.16   &0.16    &0.16  &0.16                        \\
			\hline
		\end{tabular}
	\end{center}
	\caption{Representative choices of SUSY parameter space. All masses are in GeV.}
	\label{spectrum}
\end{table}

 In Table \ref{cutflow1} and \ref{cutflow2} we present the effects of selection of cuts in simulation for both the signal and  backgrounds respectively. In addition to the SM backgrounds which are mainly due to $t\bar t$ and $WZ$, we also simulate the background taking into account the contributions due to flavour conserving decay of $\chi_2^0$ for each of the representative points in Table \ref{spectrum}.
 There are other sub-dominant backgrounds like $tbW$, $ZZ$ if one lepton is missed or $WW$, if jets fake as leptons. However these backgrounds are expected to be very small and not considered here.
 We present results for signals corresponding to those representative parameter space as shown in Table \ref{cutflow1} . In this table, the first column shows the sequence of cuts applied in the simulation, while the second column onwards event yields for the signal are shown. Table \ref{cutflow2} presents the same for the backgrounds due to SUSY in the second column and the SM in the third column. Notice that lepton isolation requirement and a cut on $\slashed{p_T}$ has considerable impact in reducing $t\bar t$ and $WZ$ background. As noted earlier, we find the SFSS criteria to be very effective in isolating the SUSY background due to flavour conserving decay of $\chi_2^0$ for all the representative points in Table \ref{spectrum} .
 Finally, it is possible to have large number of tri-lepton events in background processes, but imposition of specific choices like OFOS and SFSS along with large missing energy cut help in isolating it to a great extent as shown in Table \ref{cutflow2} . 
   In spite of this suppression of background events, the signal yields are far below than the total background contribution owing to it's huge production cross sections as shown in Table \ref{cutflow2}. Therefore, in order to improve signal sensitivity further, we impose additional requirements by looking into the other characteristics of signal events. For example, signal events are free from any kind of hadronic activities at the parton level \text{i.e.} no hard jets are expected in the signal final state, whereas in background process, in particular events from $t\bar t$ are accompanied with large number of jets. We exploit this fact to increase signal sensitivity by adding following criteria.

\begin{table}
	\begin{center}
		\begin{tabular}{|l|cccccc|}
			\hline
			&\multicolumn{6}{|c|}{Signal($\chi_2^0\chi_1^\pm$)} \\
			\cline{2-7}
			\hline 
			\multicolumn{1}{|c|}{$M_2\Longrightarrow$} &200&300&400&500&600&700\\
			\hline
			No. of events generated                     &10000  &10000 &10000 &10000   &10000  &10000      \\
			$p^{\ell_{1,2}}_{T}>20,p^{\ell_3}_{T}>10,|\eta|<2.5 $          &1371   &1752  &2014  &2218   &2225  &2342     \\         
			Lepton isolation cut                        &1330   &1669  &1883  &2055   &2036  &2112       \\
			$\slashed{p_{T}}>100 $                      &474    &959   &1326  &1600   &1683  &1860               \\
			OFOS                                        &470    &952   &1319  &1581   &1659  &1828                  \\
			Z mass veto                         &423    &849   &1218  &1485   &1574   &1752             \\
			SFSS                                        &223    &462   &640   &783   &804   &892             \\
			\hline
		
			Case a:	 jet veto                          &91     &205   &288   &337   &346    &380    \\
			Case b:	 $b$-like jet veto                         &221    &458   &635   &777   &798    &884       \\
			Case c:	 $n_j\leq 1$  and $b$-like veto         &161    &375   &479   &604   &617    &687       \\
			
			\hline

		\end{tabular}
	\end{center}
	\caption{Event summary for  signal after all selections. All energy units are in GeV.  }
	\label{cutflow1}
\end{table}

\begin{table}
	\begin{center}
		\begin{tabular}{|l|cccccc|cc|}
			\hline
			&\multicolumn{6}{|c|}{SUSY($\chi_2^0\chi_1^\pm$)}&\multicolumn{2}{c|}{SM} \\
			\cline{2-9}
                                  & A    & B     & C     &D      &E        &F   & $t\bar{t}$         & WZ   \\  
			\hline
			\multicolumn{1}{|c|}{$M_2\Longrightarrow$} &200&300&400&500&600&700&-&-\\
			\hline
			Cross section ($fb$) at 14 TeV       &$1.65\times 10^3$&370.5 &118.8&45.6 &20.5  &9.57  &$9.3\times 10^5$   &$4.47 \times 10^4$     \\
			\hline
			No. of events generated                     &10000  &10000 &10000 &10000   &10000  &10000   &$10^7$   &$3\times10^6$   \\
			$p^{\ell_{1,2}}_{T}>20,p^{\ell_3}_{T}>10,|\eta|<2.5 $   &1299   &1779  &2015  &2195   &2245    &2361    &164895   &23960 \\         
			Lepton isolation cut                        &1251   &1672  &1874  &2044   &2051   &2131     &70233    &22366   \\
			$\slashed{p_{T}}>100 $                      &454    &967   &1311  &1624   &1722   &1872     &19241    &1669            \\
			OFOS                                        &209    &482   &656  &820     &855    &918      &14012    &858              \\
			Z mass veto                                 &126    &346   &547  &728     &768    &853      &12395    &122          \\
			SFSS                                        &4      &6     &11   &14      &15     &25       &4598     &22            \\
			\hline
		   Case a:	 jet veto                   & $\leq 1$  &1   &1      &5       &4      &4        &29       &$\leq$ 1  \\
			 Case b:	 $b$-like jet veto               &4       &5   &10     &14       &13     &23       &131      &13      \\
			 Case c:	 $n_j\leq 1$ and $b$-like veto   &1       &3   &7      &9       &9      &19        &48       &5      \\
			
			\hline

		\end{tabular}
	\end{center}
	\caption{Event summary for SUSY and SM background. All energy units are in GeV. }
	\label{cutflow2}
\end{table}
 
Case a: Jet veto\\
In this case we reject events if it contain any hard jets.
In Table \ref{cutflow2} we see that while the jet veto criteria reduces the $t\bar{t}$ and WZ background significantly, but it also substantially damage the signal by a factor of 2 or 3 as shown in Table \ref{cutflow1} .
In signal process, jets arise mainly from the hadronic radiation in initial and final states and it is true for all the representative signal points.
The reason can be attributed to enhancement of hadronic activities at higher energies. 
Nevertheless the jet veto seems to be useful to improve signal to background ratio.
 However we consider two more alternatives with a goal to increase signal sensitivity further:

Case b: $b$-like jet veto\\
Here we eliminate events if there be at least one $b$ like jet.
 As can be seen Table \ref{cutflow2}, $b$ jet veto is more efficient than the jet veto condition, as the $t\bar t$ background is suppressed by a few orders of magnitude without costing the signal too much.
   
Case c: Apply $b$-like jet veto and number of jets $n_j\leq 1$\\
Here we apply the $b$-like jet veto condition along with the presence of maximum one jet.
 As seen in Table \ref{cutflow2} , it is very helpful in reducing the $t\bar{t}$ background  significantly but it does not affect the signal as much as the simple jet veto condition  ($n_j=0$) does alone.

Note that we have identified $b$-like jets by a naive jet-quark 
matching which is an overestimation from the realistic b-jet tagging\cite{Chatrchyan:2012jua}
which is out of scope of the present analysis. 
However, for the sake of illustration, we present these results with b-jet veto, (case (b) and (c)),
to demonstrate that this criteria might be very useful
in suppressing backgrounds, which requires more detector based 
simulation. In view of this, we focus only on the results obtained 
by using jet veto, case(a) for further discussion.

\begin{figure}
	\begin{tabular}{cc}
		\includegraphics[width=9.2cm]{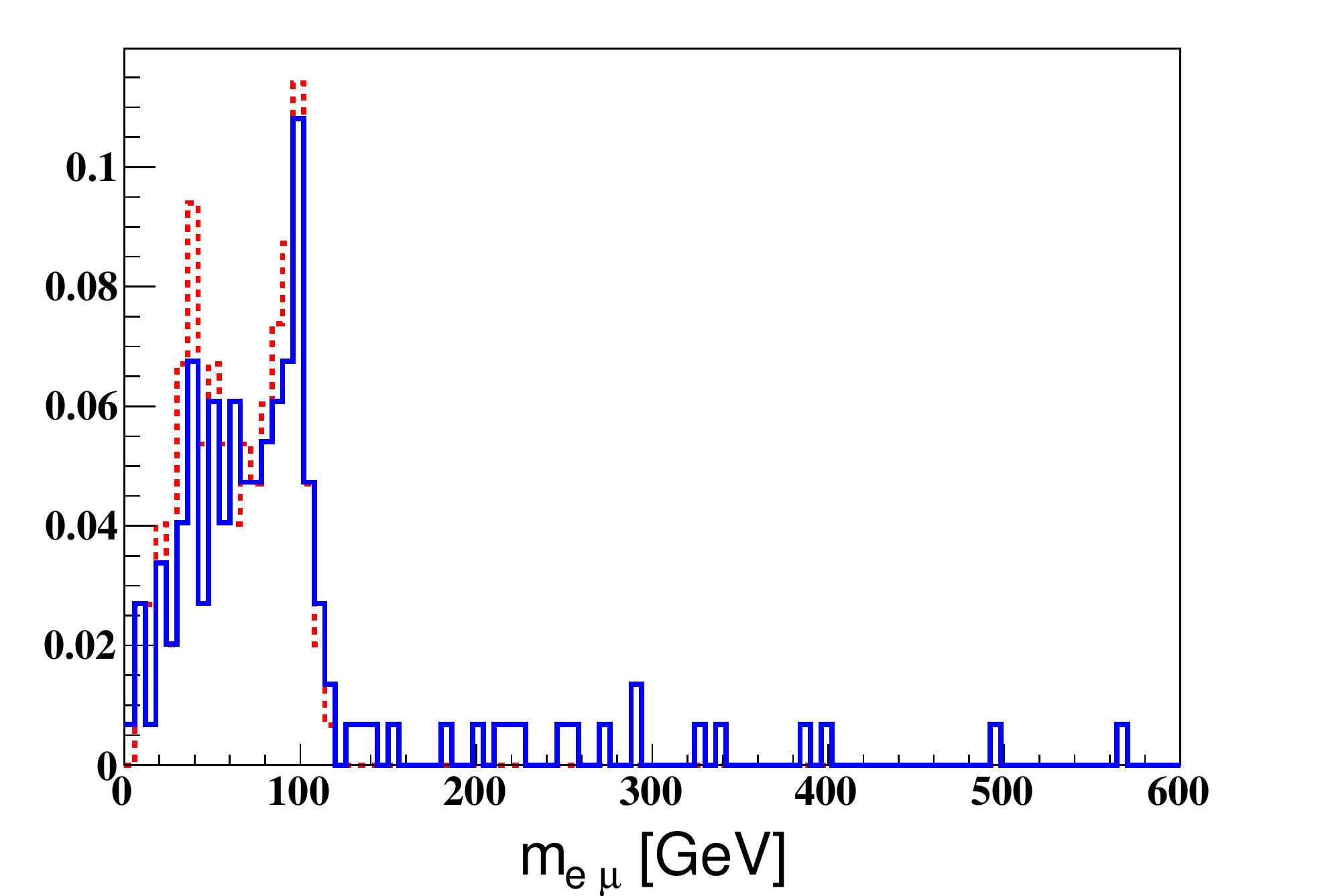}&\includegraphics[width=9.2cm]{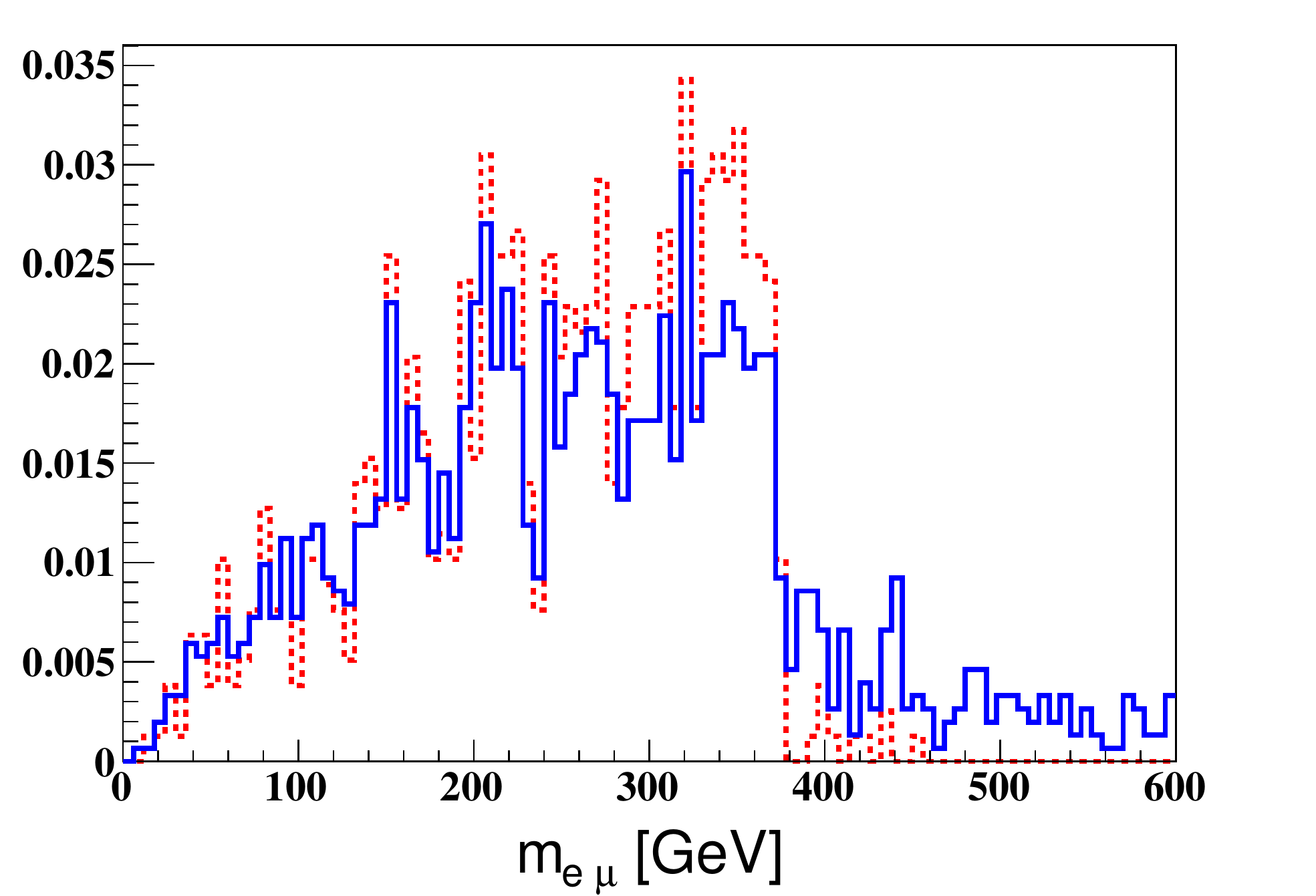}\\
	\end{tabular}
	\caption{OFOS dilepton invariant mass distribution for spectrum A (left) and spectrum F (right). The events are selected at the SFSS level.}
	\label{dilepinvmass}
\end{figure}

We also present the dilepton $(e\mu)$ invariant mass distributions for the spectrum A (left) and F (right) in Fig.\ref{dilepinvmass} normalizing it to unity. It is subject to all primary selection cuts on leptons and jets, including the OFOS and SFSS combination. The $m_{e\mu}$ distribution is expected to have a sharp edge on higher side, which can be derived analytically from kinematical consideration. The position of this edge of  $m_{e\mu}$ is given as \cite{Hisano:2002iy,Gjelsten:2004ki}, 
  \begin{eqnarray}
 (m^{max}_{e\mu})^2=m_{\chi_2^0}^2\left(1-\frac{m_L^2}{m_{\chi_2^0}^2}\right)\left(1-\frac{m_{\chi_1^0}^2}{m_L^2}\right).
  \end{eqnarray}
  The appearance of an edge in the  $m_{e\mu}$ distribution is a clear indication of LFV vertex in the $\chi^0_2$ decay. However, this  $m_{e\mu}$ distribution is affected by a combinatorial problem. For each tri-lepton event, two OFOS pairs can be constructed:
a) both leptons  coming from $\chi^{0}_2$ decay and b) ``imposter" pair  with one lepton from $\chi^{0}_2$ and the other from $\chi^\pm_1$. In Fig.\ref{dilepinvmass} the red (dotted) curve represents the dilepton invariant mass distribution of the leptons tracked to the $\chi^0_2$ vertex while blue (solid) curve corresponds to dilepton without any prior information about their origin. It (red dotted line) exhibits a very distinct edge as the identity of the lepton pair originating for $\chi^0_2$ is known a-priori. The (solid) blue  line is more realistic as it includes both the correct OFOS and SFSS pair as well as the contamination due to the ``impostor" pair which is responsible for a tail beyond the edge.
As a result it exhibits a more diffused behaviour near the position of the edge. However, we can roughly estimate the position of the edge using the blue (solid) line as $\sim 120$ GeV for the left panel and $\sim 375$ GeV for the right panel. We find that these values are in fairly good agreement with the corresponding numbers used in our simulation.
It may be noted here that such distributions with a sharp edge are the characteristic feature of these type of decays which can also be exploited to suppress backgrounds \cite{Hisano:2002iy} in order to increase signal to background ratio.

		\begin{table}
	\begin{center}
		\begin{tabular}{|l|cccccc|cc|}
				
			\hline
			
		&	\multicolumn{6}{|c|}{Signal (S)} &\multicolumn{2}{c|}{ Background (B)}\\ 
			\cline{2-9}
			
\multicolumn{1}{|c|}{Properties}                                        &A    & B   & C     &D     &E    &F   & $t\bar{t}$         & WZ   \\  
\hline
\multicolumn{1}{|c|}{Cross section ($fb$) at 14 TeV}        &$1.65\times 10^3$&370.5 &118.8&45.6 &20.5  &9.57  &$9.3\times 10^5$   &$4.47 \times 10^4$     \\
			\hline
\multicolumn{9}{|c|}{Normalized cross sections}\\
			\hline
			Case a: jet veto    &15.01   &7.59   &3.41   &1.51   &0.67    &0.37  &2.69          &$\leq 1$  \\
			Case b: $b$-like veto                         &36.4    &16.9   &7.54   &3.54   &1.63    &0.85 &12.1          &0.19      \\
			Case c: $n_{j}\leq$ 1 and $b$-like veto         &26.5    &13.9   &5.7   &2.75   &1.26    &0.66 &4.4           &0.07     \\

			\hline
			\multicolumn{9}{|c|}{$\frac{S}{\sqrt{B}}(@100)$ fb$^{-1}$}\\
			\hline
			Case a: jet veto   &91.43   &45.93  &20.78     &9.32   &4.31   &2.24   &-            &-\\	
			Case b: $b$-like veto  &100.99&47.87&21.34 &10.04  &4.64  &2.43   &-            &-\\
			Case c: $n_{j}\leq$ 1 and $b$-like veto  &122.4&64.4&26.4 &12.8  &5.92  &3.12   &-            &-\\

			\hline
		\end{tabular}
	\end{center}
	\caption{Normalized cross-section ($fb$) and $S/\sqrt{B}$ for signal and background subject to three selection conditions}
	\label{ctfr}
\end{table}

\section{Results and Discussions}
\label{lhc}
Table \ref{ctfr} gives the normalized signal and background cross-sections  due to all selection cuts. 
These are obtained by multiplying the production cross section given in the first row by acceptance efficiencies.
The production cross section are estimated by multiplying the leading order (LO)
cross section obtained from {\tt PYTHIA8} with the corresponding 
$k$ factors \footnote{The appropriate $k$ factors for $t\bar t$ and 
	WZ processes is 1.6~\cite{Kidonakis:2011tg} and 
	1.7~\cite{Campbell:2011bn} respectively while for the signal  
	it is 1.5~\cite{Beenakker:1999xh}. }.
Corresponding to these signal and background cross sections, we also present the signal significance by computing
$S/\sqrt{\rm B}$ for integrated luminosity 100~fb$^{-1}$ as shown in the bottom of Table \ref{ctfr}. Although case(b) corresponding to $b$-like jet veto results in the largest cross section for all 
signal parameter space, signal significance does not improve  
due to comparatively less suppression of SM backgrounds.      
 With the increase of gaugino masses acceptance efficiencies goes up 
as final state particles become comparatively harder, but $S/\sqrt{B}$ is
depleted due to drop in $\chi_2^0\chi_1^\pm$ pair production cross-section.
While estimating signal rates and significance, we assume  
a maximal flavour violation \textit{i.e.} $\mathcal{B}_{LFV}=1$. 
Obviously, a further suppression is expected by a factor $\mathcal{B}_{LFV}$  
which depletes the BR of $\chi_2^0$, (see Eq.\ref{chi2decay}). For a given $\delta_{12}$,  $\mathcal{B}_{LFV}$ is a function of the slepton mass as well as the mass splitting $\Delta m$ as shown in Fig. \ref{deltammL}.
   For instance $S/\sqrt{B}$ may suffer by an order of magnitude 
for $\mathcal{B}_{LFV}=0.1$. 
While the lower end of the spectrum can lead to a larger $S/\sqrt{B}$, 
the corresponding $\mathcal{B}_{LFV}$ decreases as we move further 
towards the IR part of the slepton spectrum. This can be attributed to 
stronger bounds on $\delta_{12}$ for lower slepton masses.
Though the lower mass is not yet ruled out, it is more economical to 
consider relatively heavier slepton masses as the bounds from current 
and future experiments will be relatively weaker.

\begin{figure}
	\begin{tabular}{cc}
		\includegraphics[width=8cm]{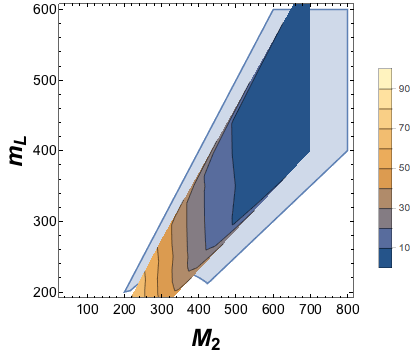}&\includegraphics[width=8cm]{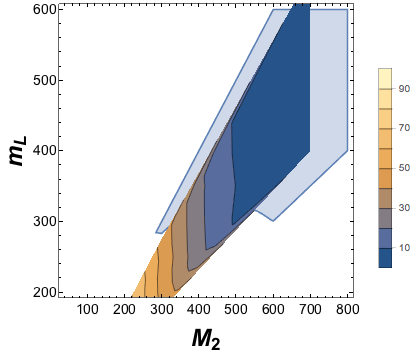}
	\end{tabular}
	\caption{Variation of $S/\sqrt{B}$ (using jet veto, case (a) for different regions with two choices of $\delta_{12}=0.01$ (left) and $\delta_{12}=0.02$(right). The regions light blue are allowed by BR($\mu\rightarrow e\gamma$) constraint.
		 Here we assume $\mathcal{B}_{LFV}=1$. Masses are in GeV.}
	\label{M2mLsb}
\end{figure}

In Fig.\ref{M2mLsb}, we illustrate this mass sensitivity by presenting $S/\sqrt{B}$ 
obtained using jet veto condition case (a).
Notice that for a given $\chi_1^\pm$ and $\chi_2^0$ masses, signal  
is not very sensitive to slepton mass as long as it is produced  
on-shell from $\chi_2^0$ decay and $M_{\chi_2^0}-m_L$ is sufficiently high.
The  regions in the $M_2-m_L$ plane correspond to different values of $S/\sqrt{B}$ computed for $\mathcal{L}=100$ fb$^{-1}$ and by
assuming $\mathcal{B}_{LFV}=1$. 
The sleptons and gaugino masses follow the parametrisation in Eq.\ref{parametrization1} and 
\ref{parametrization2}. 
It is superimposed on the region satisfying 
$BR(\mu\rightarrow e\gamma)<5.7\times 10^{-13}$  
for $\delta_{12}=0.01$ (left) and $\delta_{12}=0.02$ (right). 
As seen from Table \ref{ctfr} and Fig.~\ref{M2mLsb}, 
the signal significance is better for lower masses due to larger $\chi_2^0\chi_1^\pm$ pair production cross section. However, it suffers 
by smaller values of $\mathcal{B}_{LFV}$ corresponding to those slepton 
masses as shown in Eq.\ref{blfv} and Fig.\ref{deltammL}.

\begin{figure}
	\begin{tabular}{cc}
		\includegraphics[width=8.5cm]{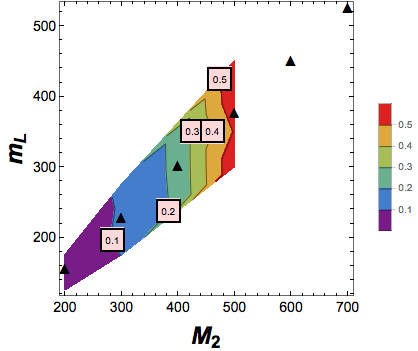}&\includegraphics[width=8.5cm]{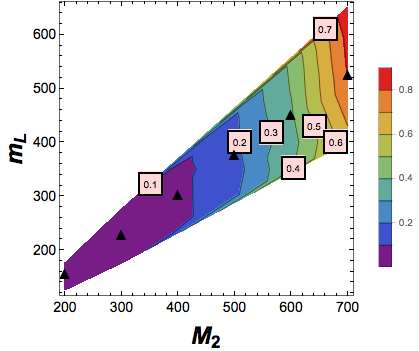}\\
	\end{tabular}
	\caption{Minimum value (in small box) of $\mathcal{B}_{LFV}$ for a $S/\sqrt{B}=5$ discovery for $\mathcal{L}=100~fb^{-1}$ (left) and $\mathcal{L}=1000~fb^{-1}$ (right).
		The $S/\sqrt{B}$ is computed using jet veto condition.	
		The filled triangles correspond  to the representative points A-F from left to right. The plot is truncated at the point where $\mathcal{B}_{LFV}>1$ is required to get a $5~\sigma$ sensitivity of signal for that particular luminosity. Masses are in GeV. }
	\label{reach}
\end{figure}

Fig.\ref{reach} shows the sensitivity reach of $\mathcal{B}_{LFV}$ in the $M_2-m_L$ plane using the parametrisation in 
Eq.\ref{parametrization1} and \ref{parametrization2}.
The numbers in boxes for different coloured regions give the 
minimum values of $\mathcal{B}_{LFV}$ which can be probed, while requiring  a $5\sigma$ discovery corresponding to those values of $M_2$  and $m_L$ and are presented for two different options of luminosities:
$\mathcal{L}=100~fb^{-1}$ (left) and $\mathcal{L}=1000~fb^{-1}$ (right). 
As the constraints from indirect flavour measurements get tighter, larger 
$\mathcal{B}_{LFV}$ can be attained with heavier slepton masses,  
while respecting bounds from the rare decays as shown in 
Fig.~\ref{parameterplot} and \ref{deltammL}. For example, for 
lower masses 	$\chi_{2}^{0}\sim\chi^\pm_1\sim 250$ GeV and 
$m_L \sim 200$ GeV, the LFV parameter $\mathcal{B}_{LFV}\sim 0.05$ or 
more can be probed at $5\sigma$ level of signal sensitivity  
for $\mathcal{L}=100~fb^{-1}$ . As expected, the minimum
$\mathcal{B}_{LFV}$ required for a $5\sigma$ sensitivity  goes up, thereby reducing the sensitivity of $\mathcal{B}_{LFV}$ measurement with the increase of gaugino and slepton masses and this can be attributed to the drop in cross-sections. 
The left plot in Fig.\ref{reach} is terminated at the point corresponding to a requirement of $B_{LFV=1}$ for a $5\sigma$ discovery. As a result, the representative 
points E and F corresponding to heavier slepton masses are beyond the 
sensitivity of LHC at $\mathcal{L}=100~fb^{-1}$ as they 
require $\mathcal{B}_{LFV}>1$ to achieve a $5\sigma$ discovery.
However, flavour violating decays with heavier slepton masses as high 
as 650 GeV can be probed with an integrated luminosity 
of $\mathcal{L}=1000~fb^{-1}$ as shown in the right plot of Fig \ref{reach}.

\section{Conclusion}
\label{conclusion}
The observation of flavour violating rare decays would be one of the best indicators of the 
existence of physics beyond the SM. 
Measurements of such decays play an important role 
in constraining several new physics models and hence has received a lot of 
attention recently. We attempt to explore the flavour violation in the lepton sector in the context of  well motivated models of flavourful supersymmetry.
We follow an approach based on a simplified model with only the left handed 
sleptons along with the neutralinos which are gaugino dominated.
We consider  pair production of $\chi^{0}_2\chi^\pm_1$ 
and their subsequent leptonic decays 
which includes the LFV decays of $\chi_2^0$. The final state is  
composed of three leptons and accompanied by large
missing energy. In addition to the presence of a lepton pair with OFOS, we observed that certain tri-lepton combinations are also characterized by a lepton pair with SFSS -which is a unique and robust signature of LFV in SUSY.

The discovery potential of observing this LFV signal is primarily 
dependent on the masses of sleptons and gauginos. These masses are however constrained
by non-observation of FCNC decays such as $\mu \to e \gamma$ and they get stronger as the flavour violating parameter $\delta_{12}$ becomes larger.
We have identified the allowed range of slepton and gaugino masses relevant for 
our study. In addition variation of LFV parameter ${\cal B}_{LFV}$
with masses of slepton and mass difference between lepton mass 
eigenstates ($\Delta m$) are also presented. 

Estimating the  various background 
contributions, we predict the signal sensitivity
for a few representative choices of SUSY parameters.  
The combination of three leptons with OFOS and SFSS is found to be very 
useful to achieve a reasonable sensitivity. It is found that for gaugino 
masses $\sim$ 250~GeV and slepton masses $\sim$200 GeV, the LFV parameter ${\cal B}_{LFV}$ as low as 0.05 can be probed with 
100~fb$^{-1}$ integrated luminosity. 
For heavier masses $\sim 600-700$ GeV, because of reduced $\chi_2^0\chi_1^\pm$ pair production cross section, the measurement of LFV parameter ${\cal B}_{LFV}$ requires higher luminosity $\sim$1000~fb$^{-1}$. 
Our study clearly establishes the prospects of finding LFV signal in this SUSY
channel at the LHC Run 2 experiment with high  luminosity options.

\section{Acknowledgements}
AI and RS would like to thank Debjyoti Bardhan and Sanmay Ganguly for discussions.

\bibliographystyle{ieeetr}

\bibliography{lfv.bib}

\end{document}